\documentstyle[aps,psfig]{revtex}
\newcommand{\eqref}[1]{(\ref{#1})}

\newcommand{\p}{{\bf p}}
\newcommand{\pf}{{\bf p}_f}
\newcommand{\vf}{{\bf v}_f}

\newcommand{\R}{{\bf R}}
\newcommand{\ber} {\begin{eqnarray}}
\newcommand{\eer} {\end{eqnarray}}

\newcommand{\tr}  {\mbox{tr}}
\newcommand{\hg}  {{\hat g}}

\newcommand{\kB}  {\mbox{$k_{\text{\tiny B}}$}}

\begin{document}

\title{Thermodynamics of a $d$-wave Superconductor Near a Surface}
\author{L.J.~Buchholtz}
\address{Department of Physics,
         California State University,
         Chico,~~Chico, CA 95929, USA}
\author{Mario Palumbo, D. Rainer}
\address{Physikalisches Institut,
         Universit\"at Bayreuth,
         D-95440 Bayreuth, Germany}
\author{J.A.~Sauls}
\address{Department of Physics \& Astronomy,
         Northwestern University,
         Evanston, IL 60208, USA }

\maketitle

\begin{abstract}

\centerline{{\bf Abstract}}
{\small\noindent
  We study the properties of an anisotropically paired superconductor
in the presence of a specularly reflecting surface.
  The bulk stable phase of the superconducting order parameter is taken
to have $d_{x^2-y^2}$ symmetry.
  Contributions by order parameter components of different symmetries
vanish in the bulk, but may enter in the vicinity of a wall.
  We calculate the self-consistent order parameter and surface free
energy within the quasiclassical formulation of superconductivity.
  We discuss, in particular, the dependence of these quantities on the
degree of order parameter mixing and the surface to lattice orientation.
  Knowledge of the thermodynamically stable order parameter near a surface
is a necessary precondition for calculating measurable surface properties
which we present in a companion paper.

 \medskip\noindent
 {\it To appear in J.~Low Temp.~Phys., Vol.~101, Dec., 1995}}
\end{abstract}


\medskip

\section{Introduction}

  Anisotropic superconductors react sensitively to electron scattering from
impurities, interfaces and surfaces\cite{buchholtz81}.
  Scattering leads, in these cases, to pair-breaking and a modification of
the superconducting state.
  This paper is a first step in a systematic study of the effects of
surface scattering on the superconducting properties of a layered system
with $d$-wave pairing.
  The possibility of $d$-wave pairing is currently under serious
consideration as a promising candidate for the proper order parameter
symmetry in the high-T$_c$ cuprates
\cite{scalapino94,pines94,hardy93,wollman93,brawner94,tsuei94,kirtley95}.
  The presence of a $d$-wave pairing interaction results in a strongly
anisotropic superconducting order parameter, and one expects that surfaces
will exert a sizable influence on the pair condensate near the boundary
\cite{buchholtz81,ovchinnikov69,ambegaokar74,kieselmann83,zhang85,buchholtz86,kopnin86,zhang87,zhang88,buchholtz91,kopnin91,kopnin92,buchholtz93}.
  Correspondingly, the physical properties of a $d$-wave system near an
interface may be profoundly different from those of an $s$-wave superconductor.
  The distinction is primarily due to the strongly enhanced pair-breaking
effects in a $d$-wave system.
  Many standard experiments, such as tunneling measurements or
electromagnetic absorption at surfaces (in the very long mean free path
limit), are particularly sensitive to the superconducting state near the
interface.
  This suggests that experiments directed at measuring effects deriving from
the interaction of the superconducting order parameter with a surface or
interface may serve as reliable probes for the symmetry of the
superconducting order parameter.
  The optimal surfaces for studying $d$-wave symmetry are those oriented
perpendicular to the layers, and we consider this geometry exclusively here.

  Because the superconducting order parameter is itself not directly
accessible to experiments, any evidence for an unconventional symmetry of
the order parameter must be more or less indirect.
  Theory is required to extract information on the order parameter from
the experimental properties of the superconductor, and to point out measurable
effects which reflect clearly the symmetry of the underlying order parameter.
  This paper and the companion paper, [II]\cite{buchholtz95b}, discuss
specific aspects of anisotropic superconductivity near surfaces.
  We consider systems with tetragonal ($D_{4h}$) crystal symmetry, which is
a good {\sl approximate} symmetry of the cuprate superconductors, and focus
on a non-degenerate order parameter (Cooper pair wavefunction) of even
parity.
  A group theoretical classification of superconducting order\cite{volovik85}
leads, in this case, to four different types of bulk order parameters
corresponding to the four one-dimensional, even parity representations of
the $D_{4h}$ group: the $A_1$, or identity representation, the $A_2$
representation, and the $B_1$ and $B_2$ representations.
  Whichever representation leads to the largest attractive coupling constant
for Cooper pairs will determine the symmetry and transition temperature of
the bulk superconducting state.
  A surface, on the other hand, may break the $D_{4h}$ symmetry and thus
may admit local contributions from all other representations, to a degree
depending on the magnitude of their coupling constants.
  In addition, surface pair-breaking may suppress the dominant bulk
order parameter and generate surface states of a new symmetry.
  Surface pair-breaking and the mixing in of new symmetries thus leads to
measurable modifications of the superconducting properties near a surface,
which carry information on the symmetry of the superconducting order
parameter.

  We investigate both the thermodynamic and spectral properties
of a superconductor in the vicinity of a specular interface.
  We first focus on calculations of the spatial dependence of the order
parameter (gap function) and of the surface free energy as functions of the
relative surface to crystal lattice orientation.
  We take the bulk order parameter to be of $B_1$ $d$-wave symmetry as
predicted by microscopic models\cite{scalapino94,pines94}, and indicated
by various experiments.
  In the presence of a surface, we consider three main order parameter
categories: those possessing a
single $B_1$ $d$-wave component (all other coupling constants being assumed
negligible), those possessing a linear combination
of $B_1$ and $B_2$ $d$-wave components (the $A_1$ and $A_2$ coupling
constants being assumed negligible), and those possessing a linear
combination of the $A_1$, $A_2$, $B_1$, and $B_2$ components (i.e.~all
components allowed by symmetry may couple in).
  We discuss to what extent the mixing of these various symmetry components
affects the overall order parameter structure in the vicinity of a surface
as well as the corresponding effect on the surface free energy.
  We use these results in [II] to study the influence of surface
pair-breaking on the tunneling density of states.

  We perform all of our calculations within the quasiclassical formulation of
superconductivity.
  This is a fully self-consistent theory which requires the numerical
calculation of the quasiclassical propagator from which the thermodynamic
and spectral data may then be extracted.
  The formulation is justified as long as the superconducting coherence
length is large compared to the Fermi wavelength.
  This approximation is known to be excellent for traditional superconductors,
but it may be of reduced validity for the high-T$_c$ cuprates due to their
substantially shorter coherence lengths.
  In section~II we give a brief review of the quasiclassical framework, and
in section~III we present our numerical results for the gap-function and
free energy.
  A discussion and summary are contained in section~IV.

\section{Quasiclassical Theory}

  The quasiclassical formulation of
superconductivity was first stated by Eilenberger \cite{eilenberger68},
Larkin and Ovchinikov \cite{larkin68,larkin76},
and Eliashberg \cite{eliashberg71}.
  This theory is capable of describing both the equilibrium and dynamical
properties of conduction electrons in either the normal or superconducting
state and may be understood as a generalization of Landau's theory of
normal Fermi liquids.
  In this section we give a brief summary of only those equations relevant
to the current analysis.
  Readers interested in a more detailed presentation are referred to the
original papers mentioned above, and to the recent review
articles\cite{eckern81,serene83,rammer86,larkin86}.
  As our primary resource regarding the quasiclassical formulation we turn
to reference\onlinecite{serene83}, whose notation we adopt whenever possible.

  In this paper we present calculations of the {\it thermodynamic}
properties of a superconducting system in equilibrium and, correspondingly,
require only
the quasiclassical theory in its imaginary-energy (Matsubara) representation.
  The fundamental quantity in the quasiclassical formulation is the
matrix propagator for quasiparticle excitations, $\hat{g}$.
  For spin-independent systems the propagator has the simple form:
\begin{equation}
\hat{g}\left(\pf,\R;\epsilon_n\right)=
\left(
  \begin{array}{cc}
    g        &  f  \\
    \bar{f}  &  \bar{g}
  \end{array}
\right) \, ,
\end{equation}
which is a $2\times 2$ matrix in particle-hole (Nambu) space.
  The propagator $\hat{g}$ is a function of a momentum on the Fermi
surface, $\pf$, a real-space position, $\R$, and an imaginary excitation
energy, $i\epsilon_n$.
  The diagonal components of $\hat{g}$ carry information on the spectrum
of Bogoliubov quasiparticles, while the off-diagonal components are related
to the Cooper pair amplitude.
  The functions $\bar{f}$ and $\bar{g}$ are related to $f$ and $g$ by
fundamental symmetry identities\cite{serene83}.

  The central equation obeyed by the propagator $\hat{g}$ is the
quasiclassical transport equation
\ber\label{trans_equation}
\displaystyle
\left[ i\epsilon_n\hat{\tau}_3 - \hat\Delta\left(\pf,\R\right) \, , \,
       \hat{g}\left(\pf,\R;\epsilon_n\right) \right]
  + i\vf\left(\pf\right) \cdot
    \vec\nabla_{\R}\,\hat{g}\left(\pf,\R;\epsilon_n\right)=0 \, ,
\eer
supplemented by a normalization condition
\ber
\left[\hat{g}\left(\pf,\R;\epsilon_n\right)\right]^2=-\pi^2\, \hat{1} \, .
\eer
  The pairing self-energy\footnote[1]{Note that we are using the
terms pairing self-energy, order parameter, and gap function
interchangeably.}
$\hat\Delta$ is computed, within the weak coupling approximation, from the
self-consistency equation:
\ber\label{gap_equation}
\hat\Delta\left(\pf,\R\right) =
  T\sum_{\epsilon_n}^{\epsilon_{co}} \oint d\pf' \; n(\pf')
  V\left(\pf,\pf'\right) \hat{f}\left(\pf',\R;\epsilon_n\right) .
\eer
  Here $V\left(\pf,\pf'\right)$ is the pairing interaction,
which determines both $T_c$ and the symmetry of the order parameter,
and $\hat{f}\left(\pf,\R;\epsilon_n\right)$ is the off-diagonal part
of the full matrix propagator $\hat{g}\left(\pf,\R;\epsilon_n\right)$.
  The integration over $\pf$ is to be interpreted as a normalized
Fermi surface integral such that $\oint d\pf \; n(\pf)= 1$, where
$n(\pf)$ is the Fermi surface anisotropy factor (see [II]).
  Equations~(\ref{trans_equation}) and~(\ref{gap_equation}) must be
solved self-consistently for the position dependent gap function
$\hat\Delta\left(\pf,\R\right)$.

\begin{figure}
\centerline{\psfig{figure=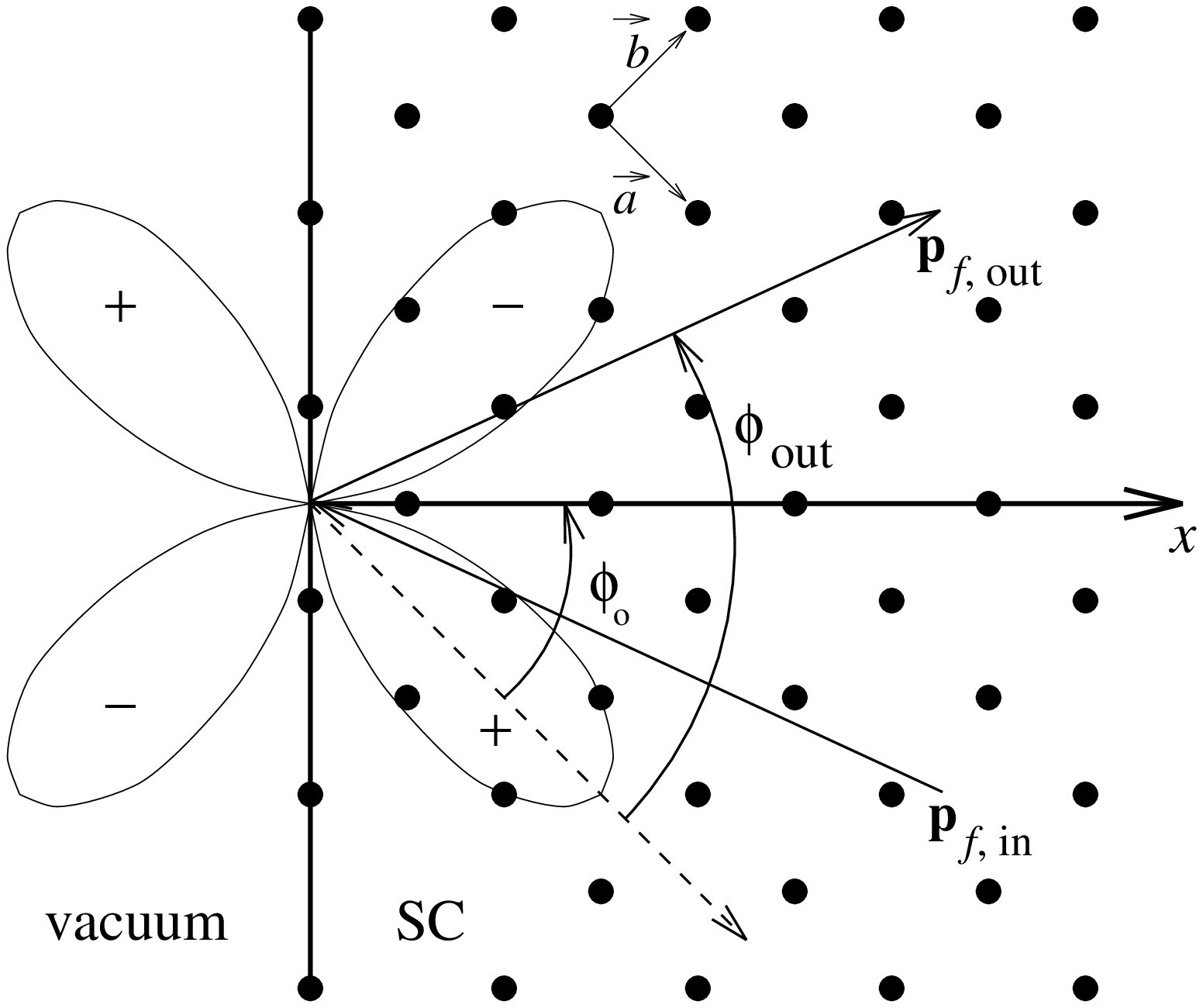,height=2.3in}}
\begin{quote}
\small
Fig.~1.~~
  This figure defines the geometrical configuration and the central
descriptive parameters of our calculation.
  We sketch the clover-shaped $B_1$ $d$-wave gap structure whose bulk
orientation is fixed to the crystal lattice.
  The angle $\phi_o$ is the surface to lattice orientation angle
($\phi_o=45\deg$ in this case), while $\phi_{\mbox{out}}$ defines the
trajectory of the out-going quasiparticle.
  The trajectory of the incoming quasiparticle is specified by
$(\phi_{\mbox{in}}-\phi_o) = \pi - (\phi_{\mbox{out}}-\phi_o)$.
\end{quote}
\end{figure}

  We consider a semi-infinite system occupying the space $x>0$ with a
single specularly reflecting interface located at $x=0$; this configuration
is depicted in Fig.~1.
  The boundary condition for the quasiclassical propagator $\hg$ at a
specular wall requires continuity along a reflected
trajectory:\cite{serene83}
\ber \label{boundary_cond}
\hg(\p_{f,\mbox{in}},\R_{\mbox{wall}};\epsilon_n) =
\hg(\p_{f,\mbox{out}},\R_{\mbox{wall}};\epsilon_n) ,
\eer
where the out-going momentum vector, $\p_{f,\mbox{out}}$, is given in terms
of the in-coming momentum vector, $\p_{f,\mbox{in}}$, and the surface
normal, $\hat{n}$, by
\ber
  \p_{f,\mbox{out}}=\p_{f,\mbox{in}}
    -2\hat{n}\left(\hat{n}\cdot\p_{f,\mbox{in}}\right) .
\eer
  Note that an anisotropic order parameter need not, in general, be
continuous at the surface along a given trajectory since the momentum
vector abruptly changes direction there.

  In our model we assume a homogeneous order parameter in the plane of
the surface; in this way there is only one relevant spatial degree of
freedom which allows us, in the following, to replace $\R$ by $x$.
  Further, we consider layered superconductors with negligibly small
interlayer coupling.
  We model these systems by a cylindrical Fermi surface with no dependence
of the gap function on $p_z$ so that the Fermi surface vector $\pf$ can be
replaced simply by the azimuthal angle $\phi$.
  These simplifications allow us to rewrite the quasiclassical transport
equation in the form
\ber\label{trans_equation_2}
\displaystyle
\left[ i\epsilon_n\hat{\tau}_3 - \hat\Delta(\phi,x) \, , \,
       \hat{g}(\phi,x;\epsilon_n) \right] +
   i v_f \cos{\phi} \frac{\partial}{\partial x} \hg(\phi,x;\epsilon_n)=0 \, ,
\eer
which amounts to an ordinary first order differential equation along
a given ``trajectory'' (incoming and outgoing) which is defined by the
parameter $\phi$.
  These trajectories may be understood as ``classical'' trajectories of
particles moving with velocity $v_f$ along the direction $\phi$ and being
reflected at the surface.
  A typical trajectory is shown in Fig.~1 where we have chosen to measure
the angle $\phi$ relative to the crystal $\hat{a}$-axis.
  In our simplified
model the Fermi surface parameters are assumed isotropic in the $ab$-plane,
and the reduction to tetragonal symmetry comes about only via the pairing
interaction.
  The use of more general Fermi surface parameters would not alter the main
results of our calculations.\footnote{A significant influence of the
Fermi surface geometry is expected if the Fermi energy lies near a Van Hove
singularity (for a discussion of these topics see
Refs.~\onlinecite{combescot88,friedel89,newns92,radtke94b}).  The effect of
non-trivial Fermi surfaces will be discussed in a future publication.}

  We incorporate contributions to the pairing self-energy from the $A_1$,
$A_2$, $B_1$, and $B_2$ representations of the tetragonal group.
  Our approximate order parameter is taken to be a linear combination of
simple trigonometric functions of the angle $\phi$:\footnote{The
precise $\phi$-dependence of the order parameter is still controversial
\cite{anderson95}.  For this study we decided to parameterize the $\phi$
dependence in a straight forward way.  An alternative approach would have
been to take the order parameter from calculations in the single-exchange
graph approximation. \cite{monthoux93}.}
\ber \label{gap_function}
\hat\Delta(\phi,x)  &=&  \hat\Delta_{s}(\phi,x)
                +  \hat\Delta_{g_1}(\phi,x)
                +  \hat\Delta_{g_2}(\phi,x)
                +  \hat\Delta_{d_1}(\phi,x)
                +  \hat\Delta_{d_2}(\phi,x) \nonumber \\
                    &=&  \hat\Delta_s(x)\times1
                +  \hat\Delta_{g1}(x)\times\cos{4\phi}
                +  \hat\Delta_{g2}(x)\times\sin{4\phi} \nonumber \\
                    & & \hspace*{19mm}
                +  \hat\Delta_{d1}(x)\times\cos{2\phi}
                +  \hat\Delta_{d2}(x)\times\sin{2\phi}
\eer
  In the usual notation, $1$ is an $s$-wave function, $\cos{4\phi}$ and
$\sin{4\phi}$ are $g$-wave functions of $A_1$ and $A_2$ symmetry
respectively, and $\cos{2\phi}$ and $\sin{2\phi}$ are $d$-wave functions of
$B_1$ and $B_2$ symmetry.
  In accordance with equation~(\ref{gap_function}), we decompose the gap
equation~(\ref{gap_equation}) into the following five self-consistency
equations:
\ber
\hat\Delta_{s}\left(x\right) &=&
  V_{s}T\sum_{\epsilon_n}^{\epsilon_{co}} \oint \frac{d\phi'}{2\pi}
  \hat{f}\left(\phi',x;\epsilon_n\right) , \label{gap_equation_s} \\
\hat\Delta_{g_1}\left(\phi,x\right) &=&
  V_{g_1}T\sum_{\epsilon_n}^{\epsilon_{co}} \oint \frac{d\phi'}{2\pi}
  2\cos{4\phi}\cos{4\phi'} \hat{f}\left(\phi',x;\epsilon_n\right) ,
  \label{gap_equation_g1} \\
\hat\Delta_{g_2}\left(\phi,x\right) &=&
  V_{g_2}T\sum_{\epsilon_n}^{\epsilon_{co}} \oint \frac{d\phi'}{2\pi}
  2\sin{4\phi}\sin{4\phi'} \hat{f}\left(\phi',x;\epsilon_n\right) ,
  \label{gap_equation_g2} \\
\hat\Delta_{d_1}\left(\phi,x\right) &=&
  V_{d_1}T\sum_{\epsilon_n}^{\epsilon_{co}} \oint \frac{d\phi'}{2\pi}
  2\cos{2\phi}\cos{2\phi'} \hat{f}\left(\phi',x;\epsilon_n\right) ,
  \label{gap_equation_d1} \\
\hat\Delta_{d_2}\left(\phi,x\right) &=&
  V_{d_2}T\sum_{\epsilon_n}^{\epsilon_{co}} \oint \frac{d\phi'}{2\pi}
  2\sin{2\phi}\sin{2\phi'} \hat{f}\left(\phi',x;\epsilon_n\right) .
  \label{gap_equation_d2}
\eer
where the $V_{X}$ are coupling constants, and the indices $s$, $d$, and
$g$ refer to $s$-wave, $d$-wave, and $g$-wave respectively.
  Note that $\hat\Delta_s(\phi,x)$ and $\hat\Delta_{g_1}(\phi,x)$ have $A_1$
symmetry, $\hat\Delta_{g_2}(\phi,x)$ has $A_2$ symmetry, and
$\hat\Delta_{d_1}(\phi,x)$ and $\hat\Delta_{d_2}(\phi,x)$ have $B_1$ and $B_2$
symmetry respectively.
  One can always eliminate the coupling constants $V_{X}$, along with the
frequency-sum cutoffs $\epsilon_{co}$, in favor of the transition temperatures
via the customary BCS relation $T_{cX}\sim1.13\epsilon_{co}e^{-1/V_X}$,
where X denotes the subscripts $s$, $g_1$, $g_2$, $d_1$, and $d_2$.
  In general, $V_X$ may be positive (attractive) or negative (repulsive).

  Equations~(\ref{trans_equation_2})--(\ref{gap_equation_d2}) must be
solved self-consistently for the spatially dependent gap amplitudes,
$\Delta_X(x)$, which are the upper right elements of the corresponding
pairing self-energies,
\ber
\hat\Delta_X(x) =
\left(
  \begin{array}{cc}
    0             &  \Delta_X(x)  \\
    -\Delta_X(x)  &  0
  \end{array}
\right) \, .
\eer
  We may then proceed to calculate various thermodynamic quantities such
as the free energy difference per unit surface area:
\ber
\Omega_{\mbox{surf}} = \int_0^\infty dx \left\{ {\cal F}(x)
              - {\cal F}_{bulk} \right\} ,
\eer
where ${\cal F}(x)$ is the free energy density in the presence of the
wall,
\begin{eqnarray}
{\cal F}(x) &=& N(E_F) \left\{
   \mbox{ln}\left(\frac{T}{T_{cs}}\right)|\Delta_s(x)|^2
 + \frac{1}{2}\left[
     \mbox{ln}\left(\frac{T}{T_{cg_1}}\right)|\Delta_{g_1}(x)|^2 +
     \mbox{ln}\left(\frac{T}{T_{cg_2}}\right)|\Delta_{g_2}(x)|^2
   \right.\right. \nonumber \\ & & \hspace*{49mm} + \left.\left.
   \mbox{ln}\left(\frac{T}{T_{cd_1}}\right)|\Delta_{d_1}(x)|^2
 + \mbox{ln}\left(\frac{T}{T_{cd_2}}\right)|\Delta_{d_2}(x)|^2
  \right]  \right\}  \nonumber \\
    \label{free_energy} &+&
  iN(E_F) \kB T \sum_{\epsilon_n>0}
  \int_{\epsilon_n}^{\infty} d\lambda \int \frac{d\phi}{2\pi}
  \left(\tr_4 \left[\hat{\tau}_3
    \left(\hg\left(\phi,x;\lambda\right) -
          \hg_0\left(\phi;\lambda\right)\right)\right] -
  2i\frac{\pi |\Delta(\phi,x)|^2}{\lambda^2}\right) .
\end{eqnarray}
and ${\cal F}_{bulk}$ is obtained by setting
$\Delta(\phi,x)=\Delta_{bulk}(\phi)$.
  Here $\hat{g}_0$ is the normal state propagator, and $\Delta(\phi,x)$ is
the full, $\phi$-dependent gap amplitude.
  In the next section we make use of the preceding equations to calculate
the various gap amplitudes and the surface free energy for various
surface orientations.

\section{Results}

  We consider a superconductor-insulator interface, with the superconductor
occupying the space $x>0$.
  We compute the constituent quantities as a function of $x$ out to a
distance of $\sim30\xi$ ($\xi=\hbar v_f / \pi\Delta_{max}(x=\infty)$) from
the wall.
  We consider an ideal interface by imposing a specular boundary
condition at the wall, which amounts to the condition:
\ber \label{boundary_cond2}
  \hg\left(\phi_{\mbox{in}},x=0;\epsilon_n\right) =
  \hg\left(\phi_{\mbox{out}},x=0;\epsilon_n\right) ,
\eer
where $\phi_{\mbox{out}}$ is the complementary angle to
$\phi_{\mbox{in}}$ (obtained upon specular reflection of the trajectory)
and is given by $(\phi_{\mbox{out}}-\phi_o)=\pi-(\phi_{\mbox{(in}}-\phi_o)$.

  We divide the presentation of our results into two subsections.
  In the first subsection we consider a simple $d$-wave order parameter
of the form $\Delta(\phi,x)=\Delta_{d_1}(x)\cos{2\phi}$,
and in the second we include the possibility of admixtures of components
of different symmetries which are induced by the interaction with the surface.

\subsection{Simple $d$-wave Model}

  In this section we consider a pairing interaction of the form
$V(\phi,\phi')=2V_{d_1}\cos{2\phi}\cos{2\phi'}$, which corresponds to a
purely $d$-wave order parameter of $B_1$ symmetry.
  This is the symmetry favored by microscopic theories to give the highest
$T_c$\cite{scalapino94,pines94}.
  The order parameter has the form:
$\hat\Delta(\phi,x)=\hat\Delta_{d_1}(x)\cos{2\phi}$ where
$\hat\Delta_{d_1}(x)$ must be evaluated self-consistently from
equations~(\ref{trans_equation_2}) and~(\ref{gap_equation_d1}).
  In Fig.~2 we plot the gap amplitude $\Delta_{d_1}(x)$ as a function of
$x$ for various surface to lattice orientation angles $\phi_o$ (the angle
$\phi_o$ is defined in Fig.~1).
  Note that for an orientation angle $\phi_o=0$ the gap amplitude is not
suppressed
by the presence of the wall.
  This is analogous to the isotropic $s$-wave case.
  However as the surface to lattice orientation angle increases the gap
amplitude rapidly becomes
suppressed in the vicinity of the wall and it completely vanishes at the
wall for $\phi_o=45\deg$.
  Note that this suppression heals on the scale of about six coherence
lengths.

\begin{figure}
\centerline{\psfig{figure=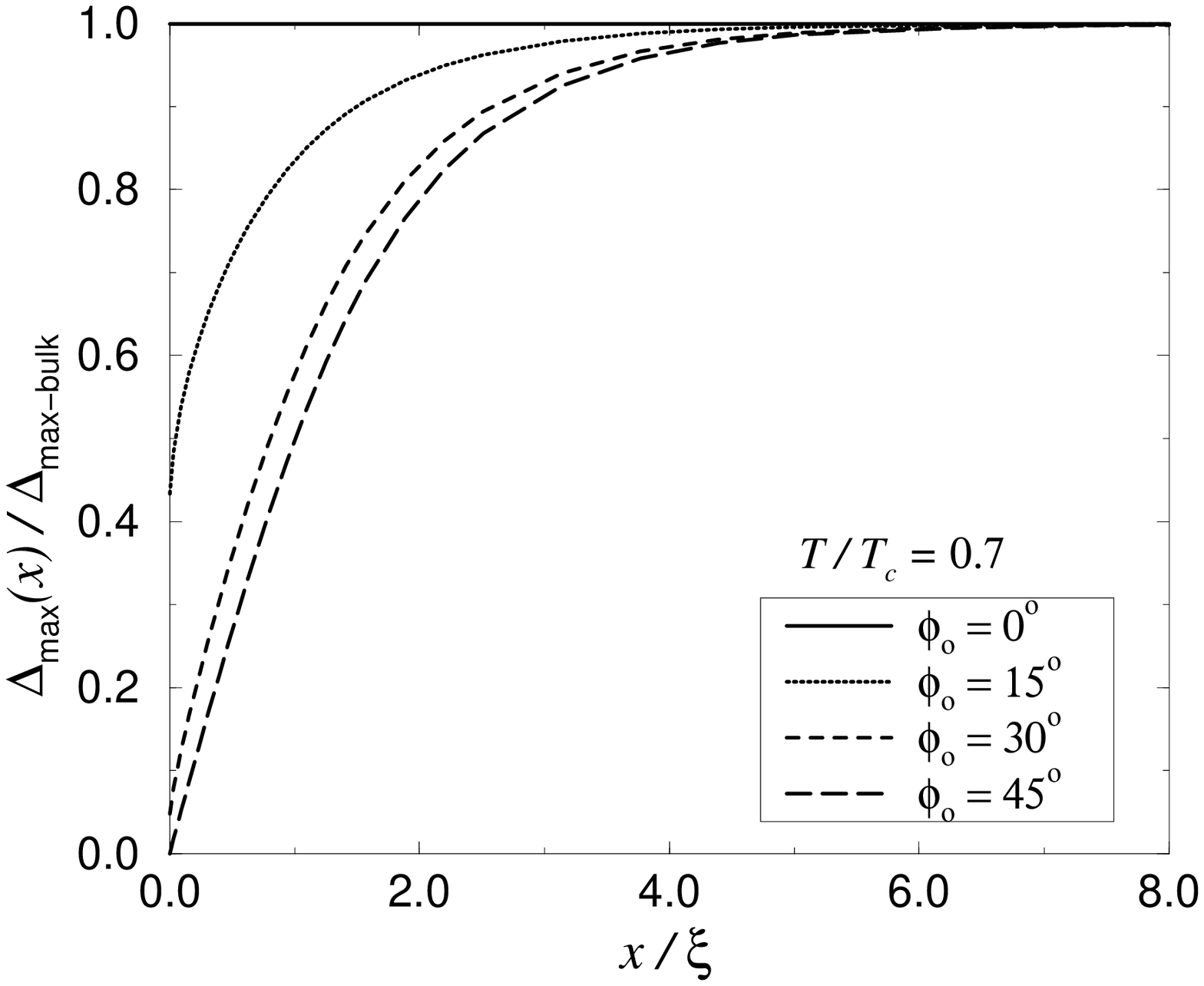,height=2.6in}}
\begin{quote}
\small
Fig.~2~~
  The order parameter amplitude as a function of position when only
one $d$-wave component is present, for various surface to lattice
orientation angles, $\phi_o$.
\end{quote}
\end{figure}

\begin{figure}
\centerline{\psfig{figure=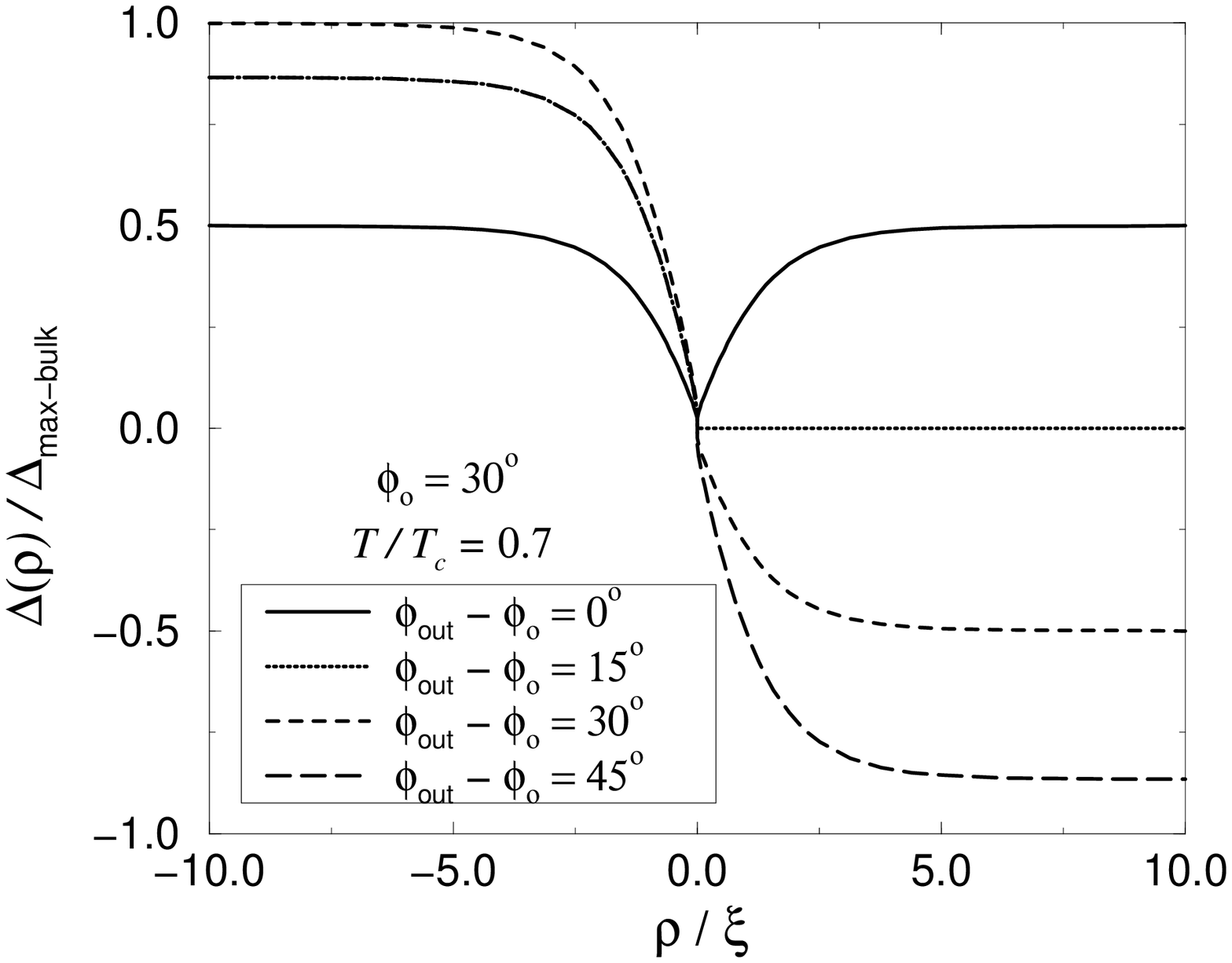,height=2.6in}}
\begin{quote}
\small
Fig.~3~~
  The local order parameter experienced by a quasiparticle moving along
a trajectory with angle $\phi_{\mbox{out}}$ for a surface to lattice
orientation angle of $\phi_o=30\deg$.
  The parameter $\rho$ is the coordinate position {\it along} the trajectory
and is related to $x$ via $\rho=x/\cos{\phi_{\mbox{out}}-\phi_o}$.
  The point of reflection is taken to be $\rho=0$.
\end{quote}
\end{figure}

\begin{figure}
\centerline{\psfig{figure=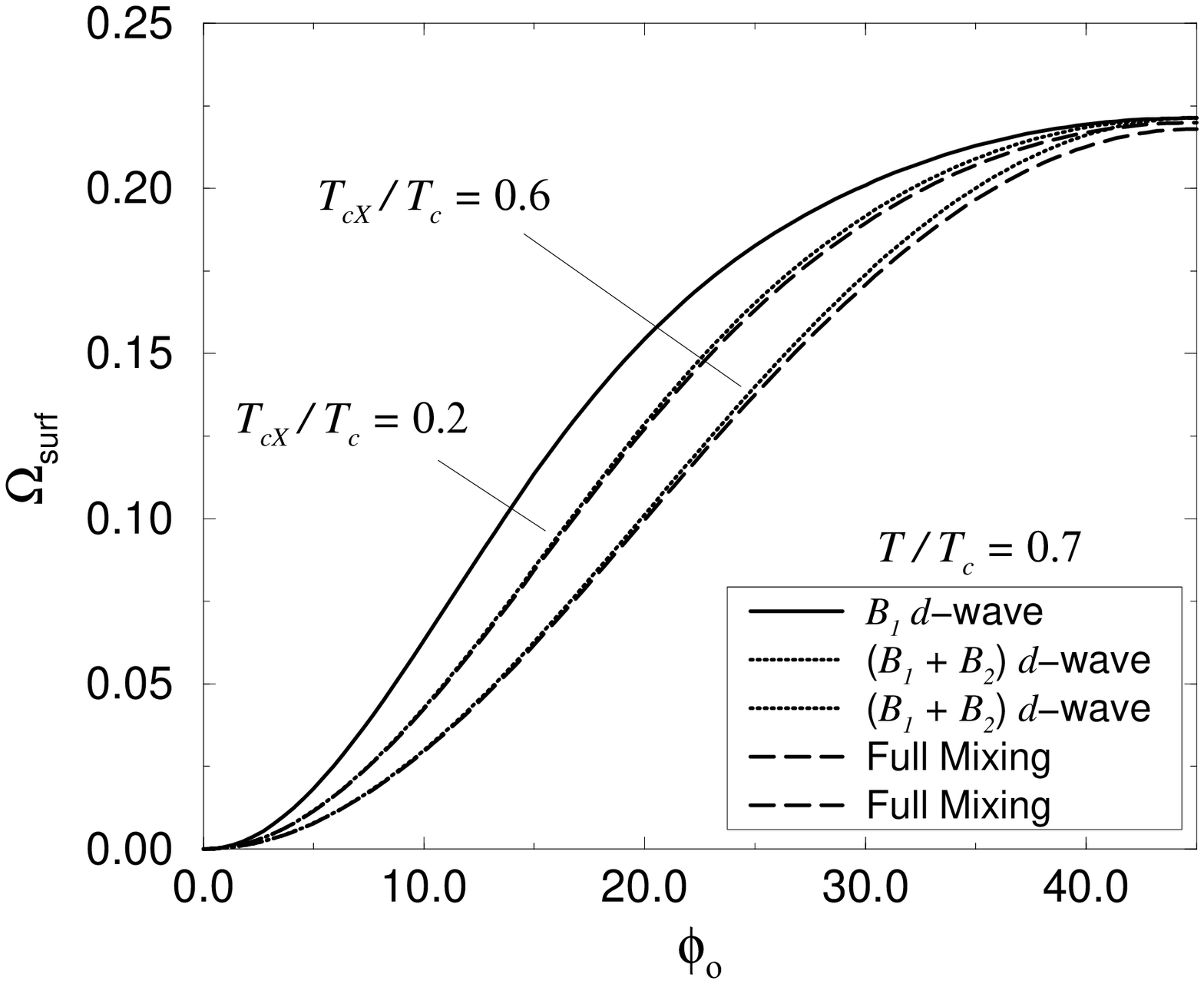,height=2.6in}}
\begin{quote}
\small
Fig.~4~~
  The surface free energy as a function of surface to lattice orientation,
$\phi_o$, for various degrees of mixing in units of
$\left(N(E_F)|\Delta_{\mbox{max-bulk}}(T=0)|^2\xi\right)$.
  The two center lines correspond to the low transition temperature
regime ($T_{cX}/T_c=0.2$) while the lower two lines correspond to the high
transition temperature regime ($T_{cX}/T_c=0.6$).
  The solid line corresponds to purely $B_1$ pairing.
\end{quote}
\end{figure}

  The angular dependence of the total gap suppression can be understood
within the framework of the quasiclassical approximation.
  A quasiparticle moving along a classical trajectory experiences the local
order parameter with momentum along $\phi$ (i.e.~in the direction of the
trajectory).
  For a surface parallel to one of the crystal axes, the local order
parameter is constant along the trajectory because
$\hat\Delta_{d_1}(\phi_{\mbox{in}})=\hat\Delta_{d_1}(\phi_{\mbox{out}})$
and the solution of the transport equation along this trajectory
is identical to the corresponding solution in the bulk.
  As a consequence, the self-consistently determined order parameter is
identical to the bulk order parameter for this orientation of the surface
and the surface free energy is zero.
  For any other surface orientation one has
$\hat\Delta_{d_1}(\phi_{\mbox{in}})\neq\hat\Delta_{d_1}(\phi_{\mbox{out}})$,
and
the propagator $\hat{f}$, and consequently the self-consistent order
parameter, will be modified by the surface.
  In Fig.~3 we plot the local order parameter $\Delta$ as a function of the
spatial position, $\rho$, {\it along} the trajectory for a surface
orientation $\phi_o=30\deg$ and several different trajectories.
  The point of reflection is taken to be at $\rho=0$, with negative
$\rho$ values denoting an incoming particle.
  This depicts the local gap amplitude that a quasiparticle experiences
along its trajectory.
  As previously mentioned, a discontinuous jump in the order parameter upon
reflection at the wall can be observed along all trajectories (except when
$\phi_{\mbox{out}}-\phi_o=0\deg$) due to the discontinuous change in
momentum direction.
  Parenthetically, we note that an asymptotic change in sign of the order
parameter along a trajectory (e.g.~Fig.~3, trajectories
$\phi_{\mbox{out}}-\phi_o=30\deg$ and $45\deg$) can give rise to significant
spectral effects such as zero-energy bound states\cite{hu94} and Tomasch
oscillations\cite{kieselmann83}.
  Our discussion of spectral effects will appear in a companion paper.

  The surface free energy for the case of a simple $d$-wave order parameter
is displayed as a function of the surface orientation angle $\phi_o$ in
Fig.~4 as
the solid line.
  The surface energy is zero, as expected, for a surface parallel to one
of the crystal axes ($\phi_o=0\deg, 90\deg$, \ldots).
  For any other orientation the surface energy is positive as a consequence
of the surface pair-breaking.
  Observe, in this case, that the energy values are very strongly dependent
on the relative orientation of the gap nodes with respect to the wall.
  These qualitative features of the pure $d$-wave model will persist even
if components of other symmetries are mixed in.
  However, as we discuss in the following subsection, mixing in leads to
significant
quantitative differences in the order parameter structure and the surface
free energy.

\begin{figure}
\centerline{\psfig{figure=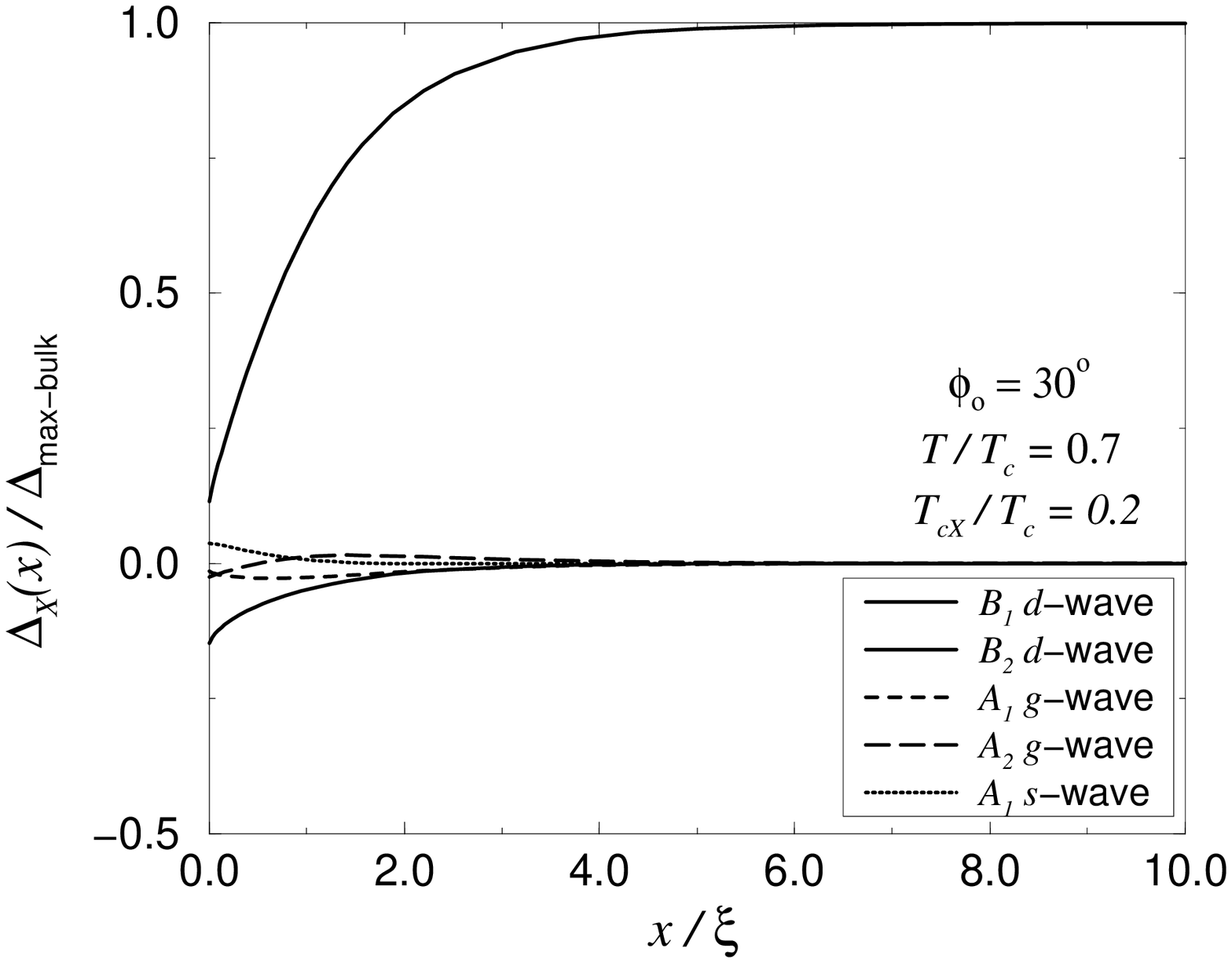,height=2.6in}}
\begin{quote}
\small
Fig.~5~~
  The individual gap amplitudes, $\Delta_X(x)$, as functions of $x$ for the
low coupling constant regime ($T_{cX}/T_c=0.2$) at a surface to lattice
orientation angle of $\phi_o=30\deg$.
\end{quote}
\end{figure}

\begin{figure}
\centerline{\psfig{figure=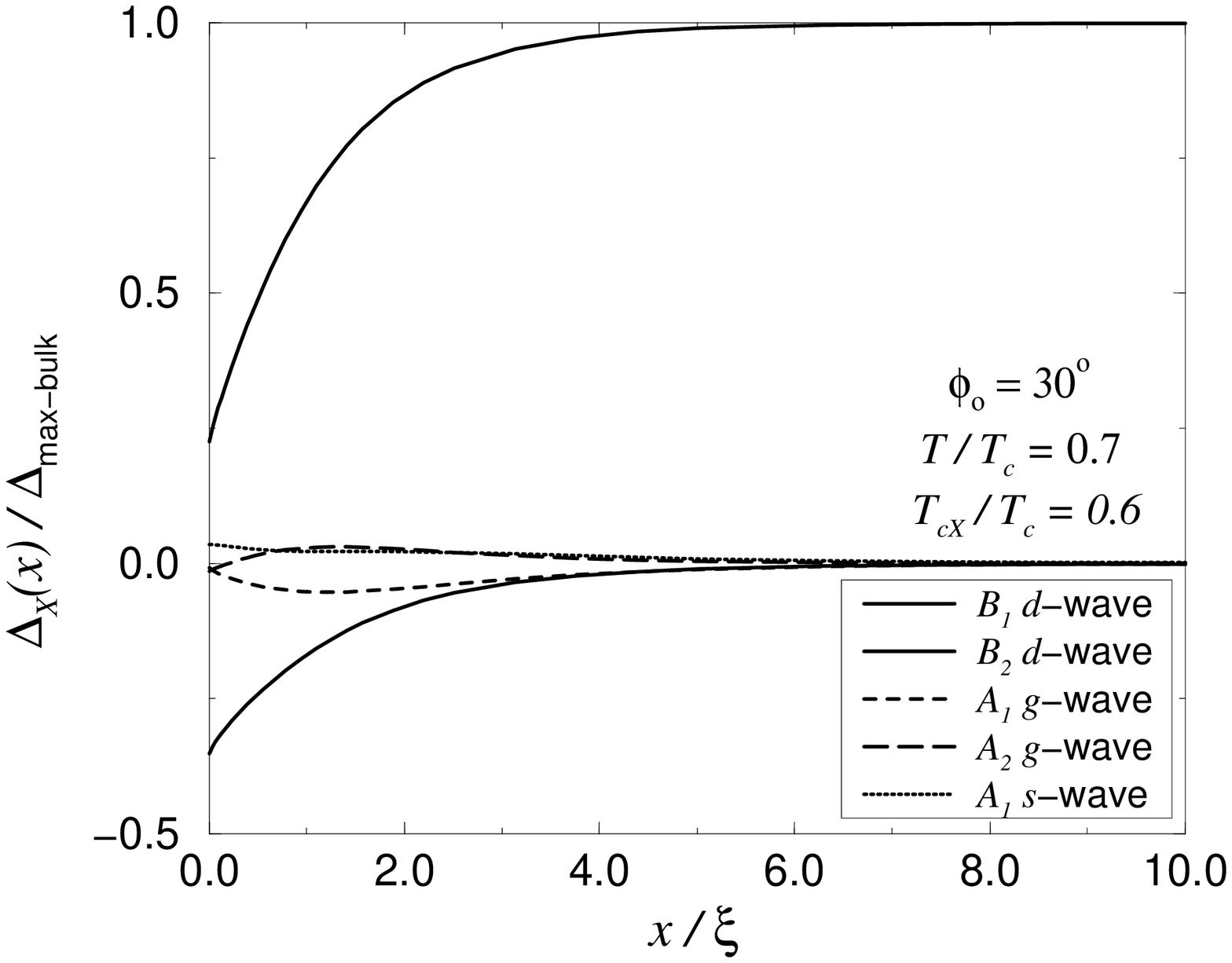,height=2.6in}}
\begin{quote}
\small
Fig.~6~~
  The individual gap amplitudes, $\Delta_X(x)$, as functions of $x$ for the
low coupling constant regime ($T_{cX}/T_c=0.6$) at a surface to lattice
orientation angle of $\phi_o=30\deg$.
\end{quote}
\end{figure}

\subsection{Extended $d$-wave Model}

  In this section we consider an admixture of the previously mentioned
components $\Delta_s$, $\Delta_{d_2}$, $\Delta_{g_1}$, and $\Delta_{g_2}$
to the dominant $\Delta_{d_1}$ component.
  Thus the total gap amplitude has the form
\ber \label{total_gap}
  \Delta(\phi,x) = \Delta_s(x) &+& \Delta_{g_1}(x)\cos{4\phi}
                                +  \Delta_{g_2}(x)\sin{4\phi} \nonumber \\
                               &+& \Delta_{d_1}(x)\cos{2\phi}
                                +  \Delta_{d_2}(x)\sin{2\phi} .
\eer
  The degree to which a component $\Delta_X$ mixes in is determined by
the corresponding transition temperature, $T_{cX}$.
  The dominant component has the highest transition temperature
($T_{cd_1}$) and provides the scale for all other temperatures in the
calculation.
  Accordingly, we have $T_c\equiv T_{cd_1}$ and then the ratios
$T_{cX}/T_{c}$ become the parameters of the calculation.
  We consider only coupling constant and temperature values for which
the non-dominant components (i.e.~$T_{cX}/T_{c}<1$) vanish in the bulk,
but may appear near the wall.
  The admixture of the subordinate components changes the momentum
dependence of the gap function.
  For instance, an admixture of $\Delta_{d_2}$ causes a rotation of the
total gap (e.g.~$\Delta_{d_2}$ alone is equivalent to a $45\deg$ rotation
in momentum space).
  The effective gap maximum and the angular position of the gap's
nodes thus vary, in general, with distance from the wall.
  By means of mixing in the new components, the system is able to alter
the gap's angular and spatial structure and thereby further lower its
free energy.

\begin{figure}
\centerline{\psfig{figure=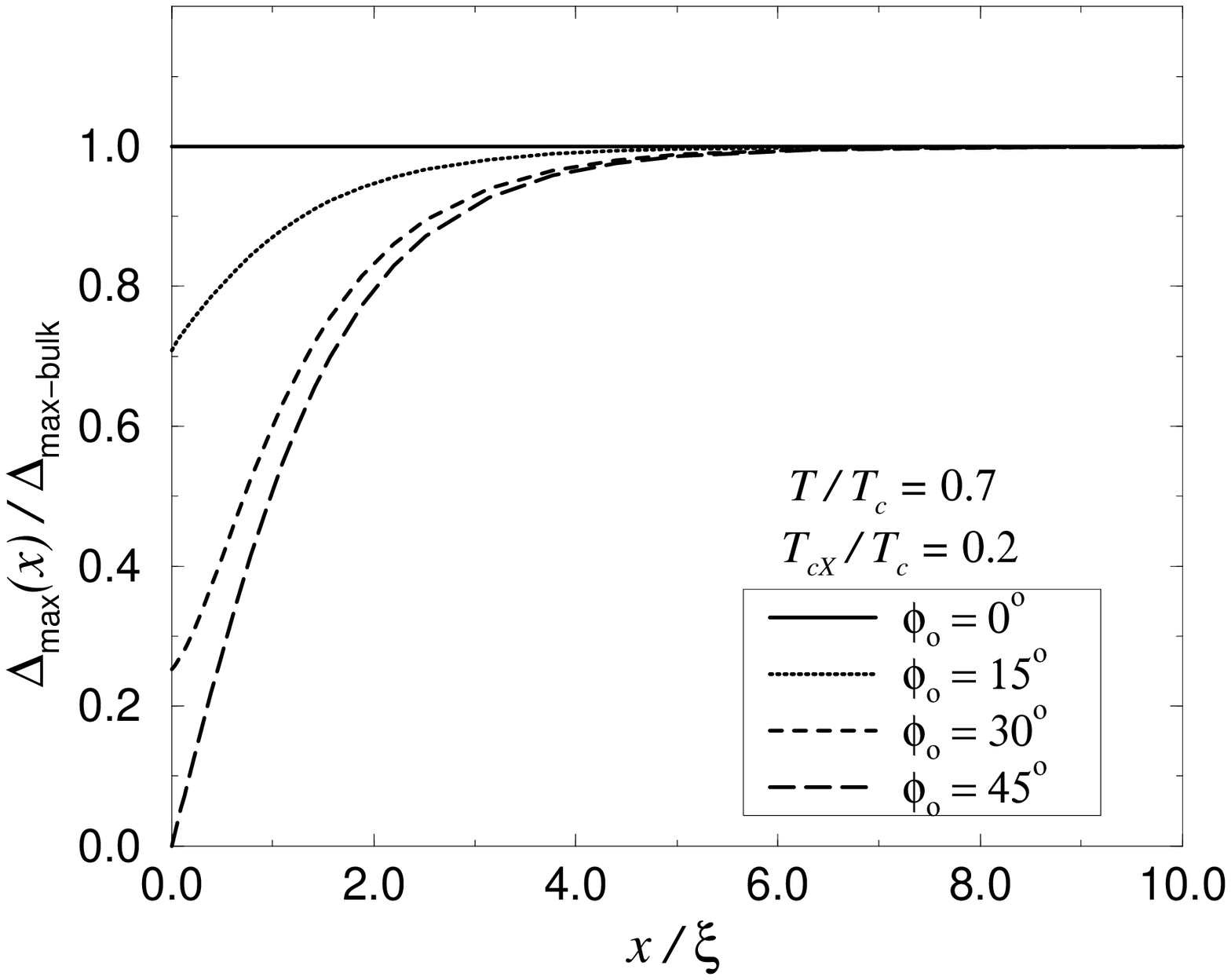,height=2.55in}}
\begin{quote}
\small
Fig.~7~~
  The maximum of the total order parameter as a function of $x$ in the
low coupling constant regime, $T_{cX}/T_c=0.2$, for various surface to
lattice orientation angles, $\phi_o$.
\end{quote}
\end{figure}

\vspace*{-0.175in}

\begin{figure}
\centerline{\psfig{figure=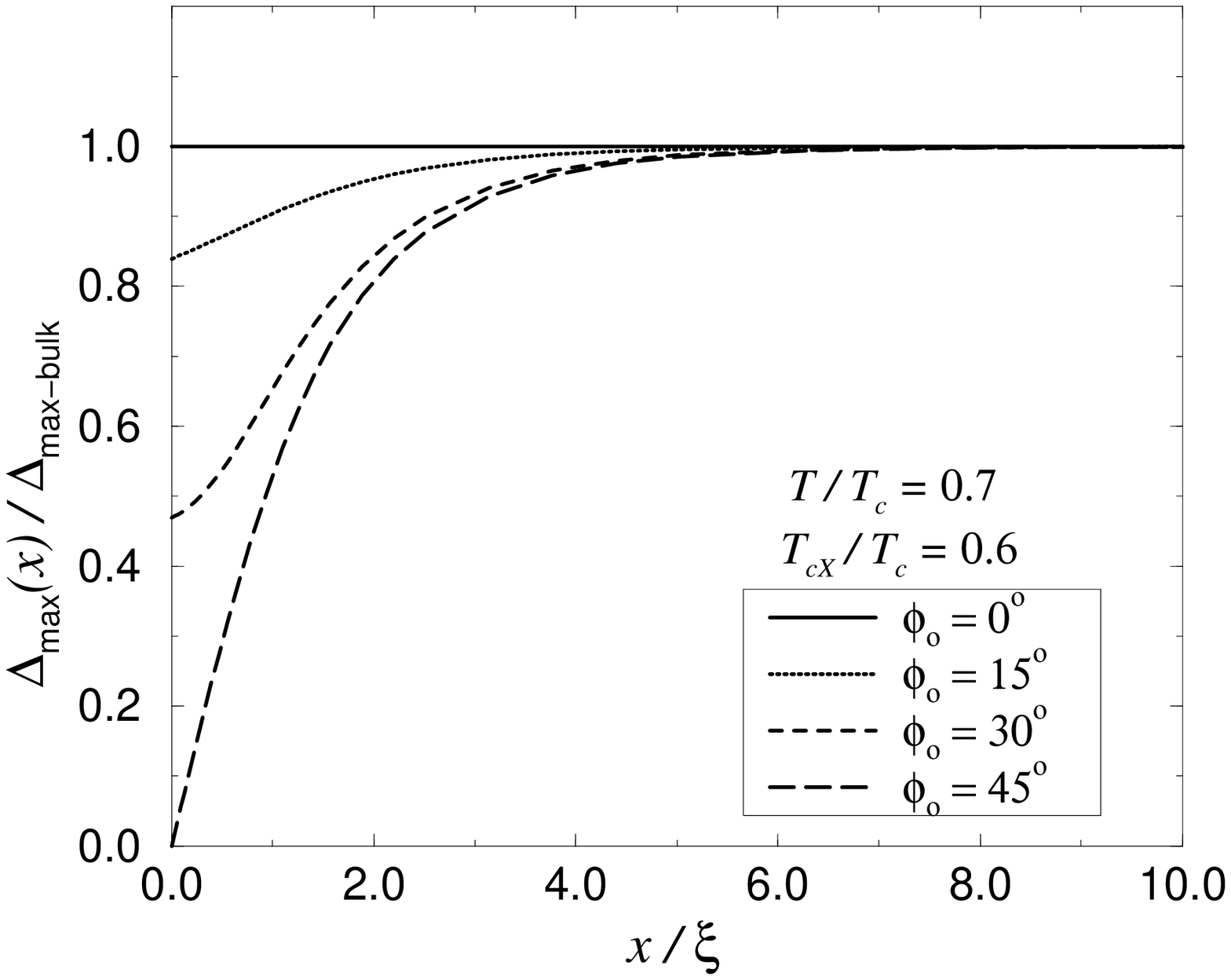,height=2.55in}}
\begin{quote}
\small
Fig.~8~~
  The maximum of the total order parameter as a function of $x$ in the
low coupling constant regime, $T_{cX}/T_c=0.6$, for various surface to
lattice orientation angles, $\phi_o$.
\end{quote}
\end{figure}

  The individual gap amplitudes are plotted versus distance from the
wall in Figs.~5 and~6 for a temperature $T/T_{c}=0.7$.
  These two graphs portray the behavior of the gap amplitudes for two
different coupling constant regimes.
  For simplicity we have set all of the admixture components to have the
same value for $T_{cX}/T_c$.
  Note that in both cases the two $d$-wave components enter with by far
the largest amplitudes and are truly the dominant representatives.
  Indeed, a computation without the $s$ and $g$-wave components left the
amplitudes of the two $d$-wave components virtually unaltered.

  The corresponding maximum gap values are plotted versus distance from the
wall in Figs.~7 and~8 and should be compared with Fig.~2 where the
admixture is absent.
  Clearly the system is now able to compensate and avoid (to some degree)
gap reduction with the level of compensation being greater for larger
$T_{cX}$.
  Note that for both transition temperature regimes the total gap maximum
for a surface orientation of $\phi_o=45\deg$ is nearly unaltered from the
corresponding curve without mixing.
  This is a special symmetric case whose physics depends strongly on
the ratio $T/T_{cX}$ and thus presents a potential mechanism for
actually measuring the admixture coupling constants.

\begin{figure}
\centerline{\psfig{figure=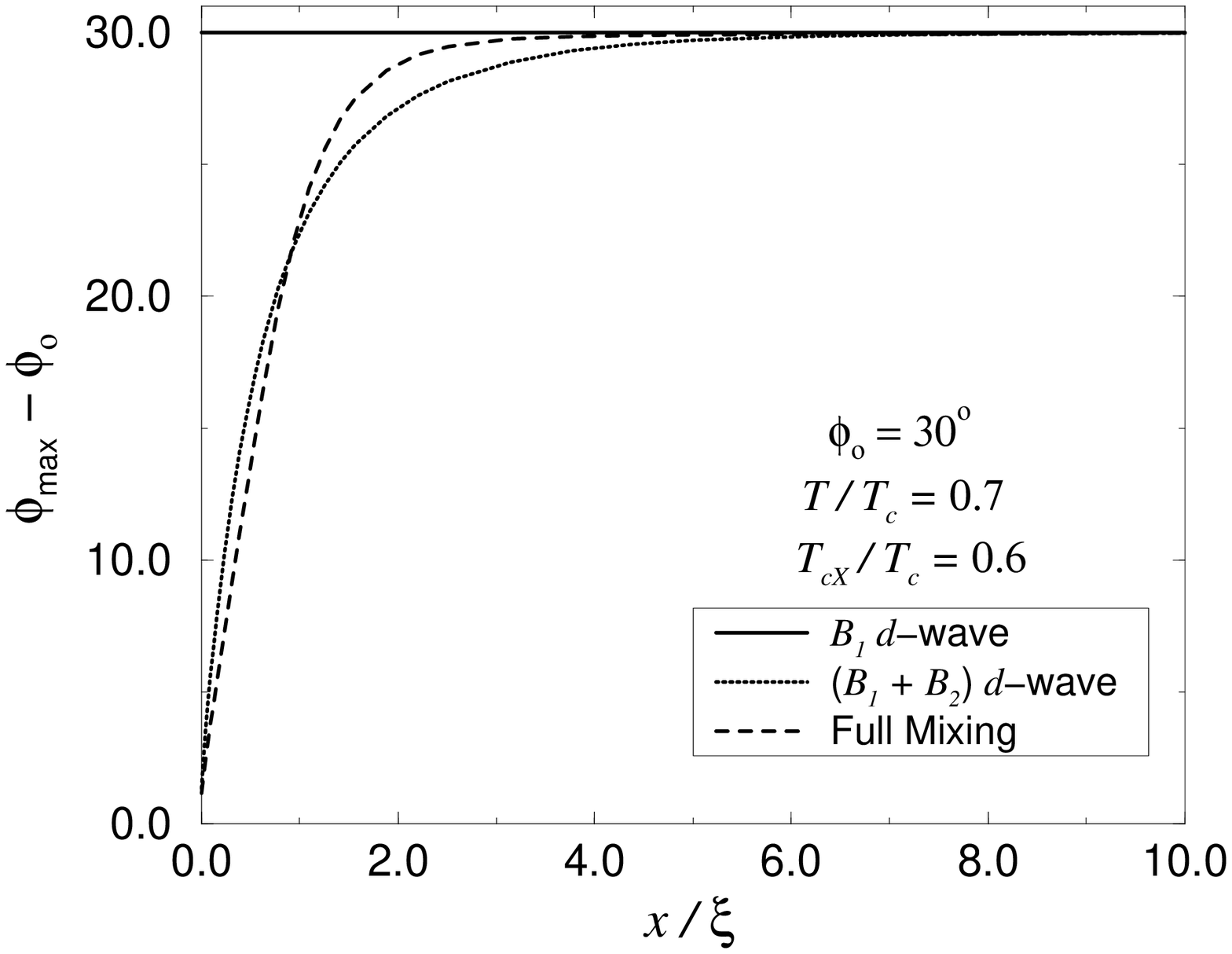,height=2.6in}}
\begin{quote}
\small
Fig.~9~~
  The angular orientation of the order parameter maximum with respect to
the surface normal as a function of $x$ for three different degrees of
mixing: a single $B_1$ $d$-wave component, a $B_1$ and a $B_2$ $d$-wave
component, and {\it full mixing}.
\end{quote}
\end{figure}

\begin{figure}
\centerline{\psfig{figure=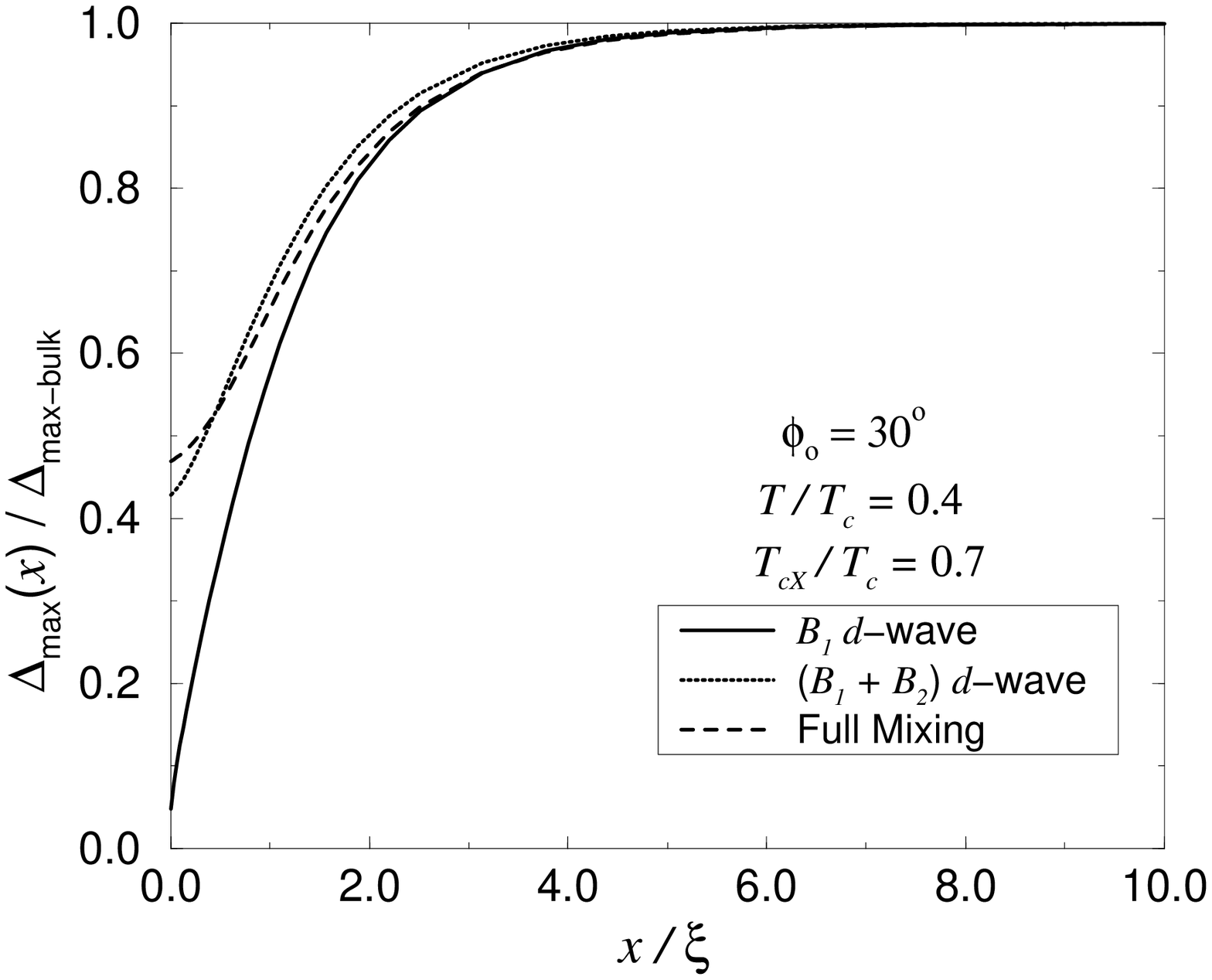,height=2.6in}}
\begin{quote}
\small
Fig.~10~~
  The maximum of the total order parameter as a function of $x$ for
three different degrees of mixing: a single $B_1$ $d$-wave component,
a $B_1$ and a $B_2$ $d$-wave component, and {\it full mixing}.
\end{quote}
\end{figure}

  Another striking quality of the self-consistent solution is the complete
reorientation of the total gap's nodes toward the optimal configuration
(i.e.~$\Delta(\phi_{\mbox{in}})=\Delta(\bar\phi_{\mbox{out}})$) at the wall,
independent of the surface
orientation and even for surprisingly small admixture coupling constants.
  This effect can be plainly seen in Fig.~9 which shows the angular
orientation of the maximum of the gap function as a function of $x$ for
different degrees of mixing.
  In the bulk $\phi_{max}=0\deg,\pm90\deg$,\ldots, however in the vicinity
of the wall the total gap rotates in an attempt to orient the maximum
gap along the surface normal ($\phi_{max}-\phi_o=0\deg,\pm90\deg$,\ldots)
at $x=0$.
  This overall rotation of the bulk $B_1$ $d$-wave gap function is achieved
by the mixing in of the $B_2$ $d$-wave component.
  The mixing in of the other components can affect the magnitude and the
shape of the total gap function, but they may not bring about an overall
rotation.
  This is presumably the reason for the system's preference of the $B_2$
$d$-wave component over the other symmetry components.

  The maximum gap values for a given surface orientation are plotted as
a function of $x$ in Fig.~10 for various degrees of mixing.
  Observe that in the immediate vicinity of the wall the addition of the
second ($B_2$) $d$-wave component has a substantial effect, while the
addition of the remaining three has a less pronounced effect.
  This interpretation is substantiated by the results of the free energy
calculations which are plotted in Fig.~4.
  We display results for a single $d$-wave component, two $d$-wave
components, and the fully mixed case.
  Clearly, the inclusion of the $A_1$ and $A_2$ components in addition
to the $B_1$ and $B_2$ components improves the free energy very little.
  Indeed, the lines lie within a few percent of each other for most surface
orientations.

\vspace*{-0.9in}
\begin{figure}
\centerline{\psfig{figure=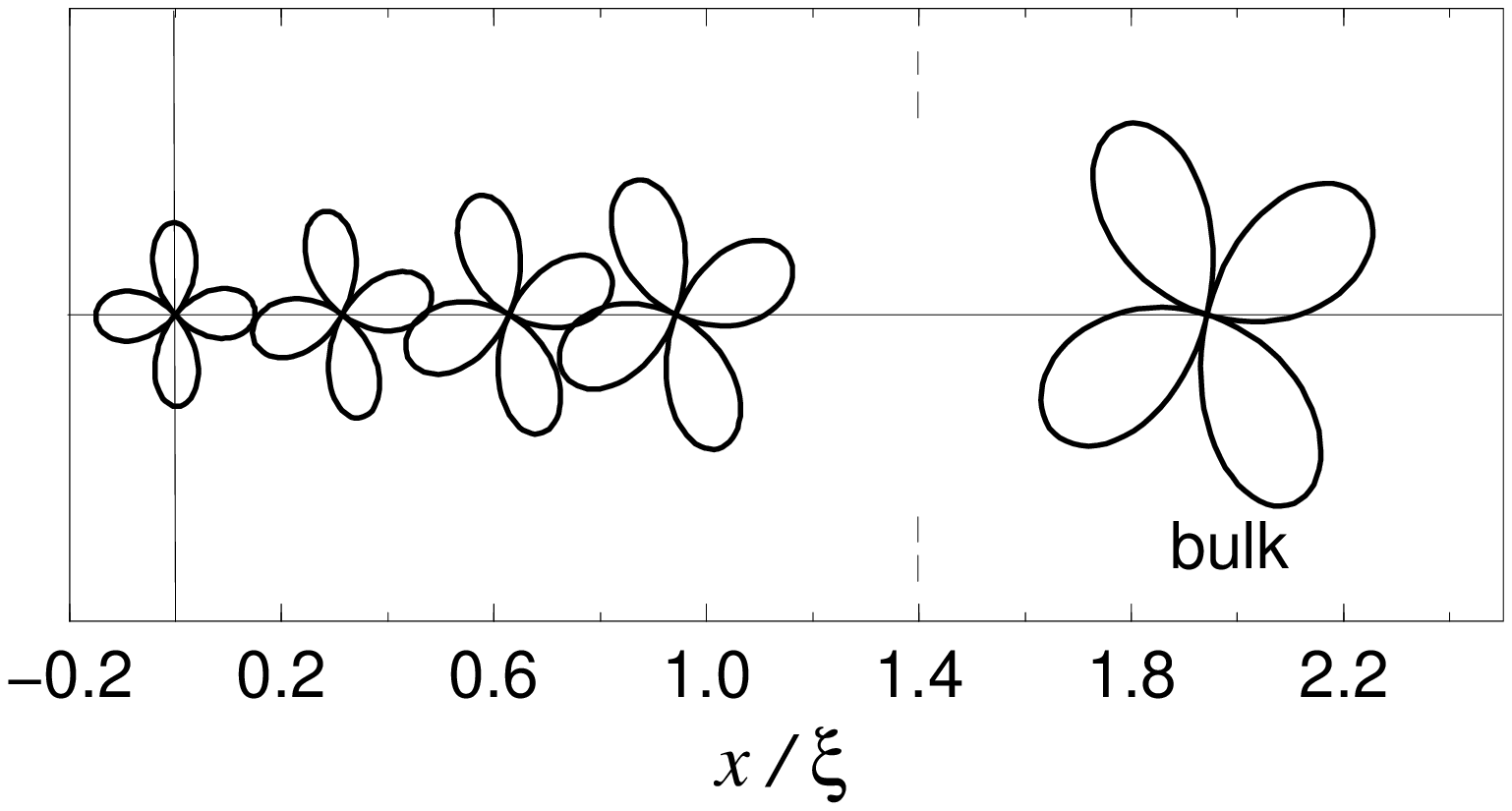,height=2.6in}}
\begin{quote}
\small
Fig.~11~~
  The orientation and magnitude of the total order parameter as a
function of distance from the wall for a surface to lattice orientation
angle of $\phi_o=30\deg$.
  The bulk order parameter (at $x=\infty$) is included for comparison.
\end{quote}
\end{figure}

\begin{figure}
\centerline{\psfig{figure=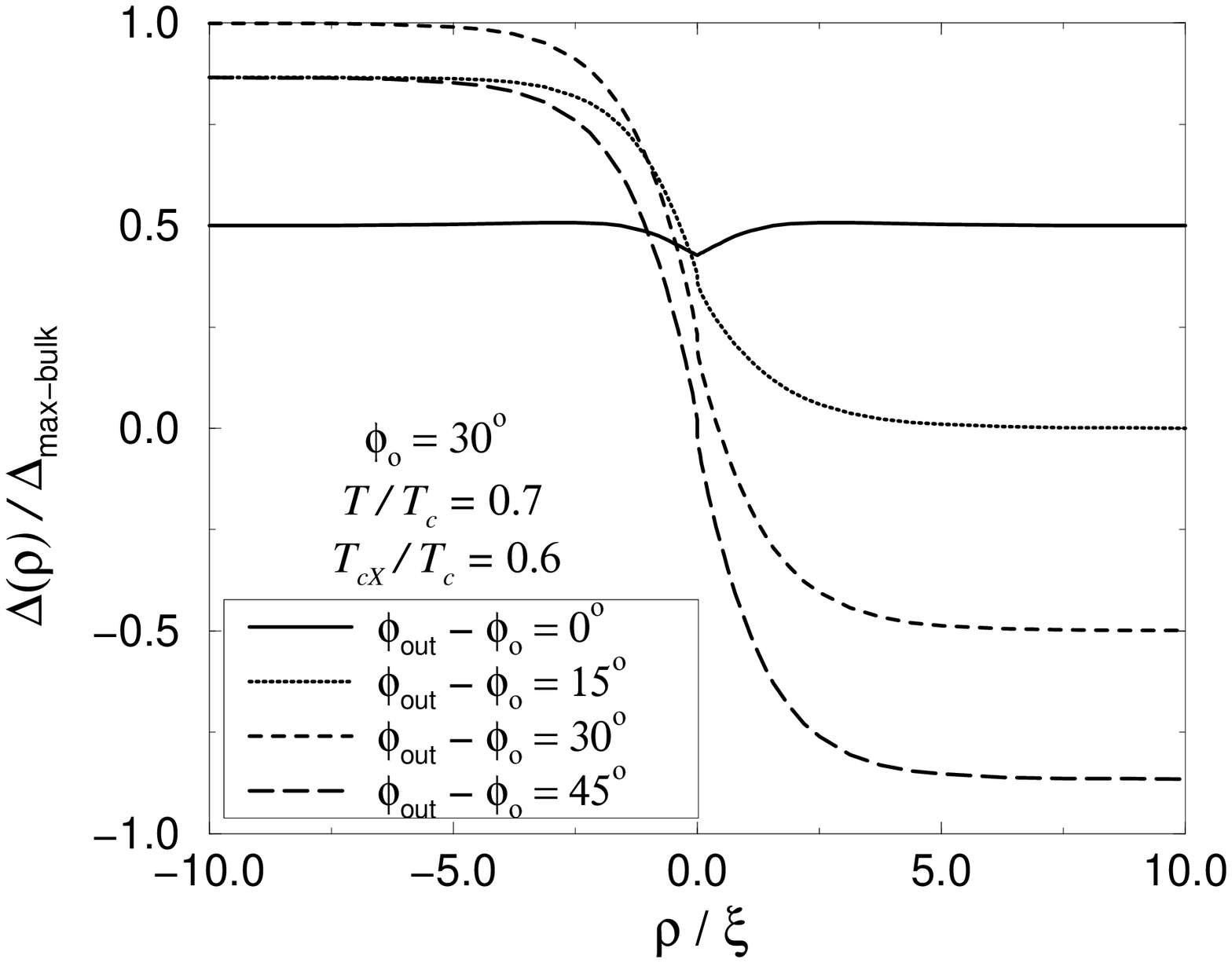,height=2.6in}}
\begin{quote}
\small
Fig.~12~~
  The local order parameter experienced by a quasiparticle moving along
a trajectory of angle $\phi_{\mbox{out}}$ for a surface to lattice
orientation angle of $\phi_o=30\deg$.
  This is the case of an order parameter consisting of $B_1$ and $B_2$
$d$-wave components.
  The parameter $\rho$ is the coordinate position {\it along} the trajectory
and is related to $x$ via $\rho=x/\cos{\phi_{\mbox{out}}-\phi_o}$.
  The point of reflection is taken to be $\rho=0$.
\end{quote}
\end{figure}

  The crucial importance of the $B_2$ components is emphasized in
Fig.~11, which depicts the angular orientation of the gap function with
respect to the surface normal as a function of position.
  In this figure we consider an order parameter consisting of only the
$B_1$ and $B_2$ $d$-wave components.
  The addition of $B_2$ allows the system to rotate the total gap
maximum toward its optimal orientation (normal to the surface) at the wall.
  This occurs for both large and small coupling constants and represents
one of the most prominent results of this calculation.
  We fully expect that this could have profound consequences for measurements
since even a small admixture assures that the order parameter assumes a
symmetric orientation at the wall {\it regardless} of its bulk orientation.

  We conclude our presentation of the numerical results by presenting a plot
of the local gap experienced by a quasiparticle along a given trajectory.
  The results for the $B_1+B_2$ mix are shown in Fig.~12 and should be
compared with Fig.~3.
  The fact that the total gap has rotated toward its optimal orientation at
the wall can be seen, for example, by examining the
$\phi_{\mbox{out}}-\phi_o=0$ trajectory where the local gap is now greater
at the wall than in the case without mixing (compare with Fig.~3).

\section{Conclusions}

  In this paper we have examined the thermodynamic properties of an
anisotropically paired superconductor in the vicinity of
a perfectly reflecting surface.
  In particular we considered an order parameter which can be expressed
as a linear combination of components transforming as the $A_1$, $A_2$,
$B_1$, and $B_2$ representations of the $D_{4h}$ (tetragonal) group.
  We carried out self-consistent calculations of the various gap
amplitudes within the quasiclassical theory of superconductivity.
  Such calculations are needed as a starting point for calculations of
a number of physically relevant quantities such as the excitation spectrum,
the surface impedance, Josephson couplings, etc.
  Another important example is the tunneling density of states which is
discussed in [II].

  Starting from the assumption that a $d$-wave order parameter transforming
as the $B_1$
representation is the bulk stable phase, we showed that even for a small
$T_{cb_2}$ the mixing in of a $B_2$ component can have a profound effect
on the order parameter structure near the wall.
  This arises from the consequent effective rotation of the total gap toward
the optimal (i.e.~symmetric) orientation in the immediate vicinity of the
wall.
  The effects of the mixing in of the $A_1$ and $A_2$ components has a
smaller,
yet noticeable, influence on the gap function, while their effect on the
surface free energy is considerably less.
  Thus the presence of the $A_1$ and $A_2$ symmetries has a minor effect
on the thermodynamic properties of the system, but can have a more
pronounced effect on the excitation spectrum.
  The strongest pair-breaking for a $B_1$ symmetry order parameter is
obtained at a $45\deg$ surface, and for a $B_2$ order parameter at a
surface along one of the crystal axis ($0\deg$ surface).
  We considered here ideal surfaces without roughness or degradation.
  This represents the optimal configuration for probing the order parameter
by surface measurements.
  Surface imperfections will wash out, to some degree, the measurable
orientational effects of an anisotropic order parameter.
  As a result, experiments should be performed preferentially at clean and
well characterized surfaces oriented perpendicular to the $ab$-planes.

\medskip
\medskip
  {\it Note Added in Proof.}
  After the submission of this manuscript we received a preprint by
M.~Matsumoto and H.~Shiba in which the possibility of the mixing in of
subdominant symmetry components which break time-reversal symmetry is
discussed.
  We have not allowed for the possibility of such states in our
calculations.
  These time-reversal symmetry breaking states are stable in certain
regions of the coupling-constant/temperature parameter space.
  All of the calculations in the current manuscript, however, were
done in a region of parameter space where these time-reversal breaking
states are not stable.

\section*{Acknowledgements}

  The research of L.J.B was supported by the Fulbright Commission and that
of M.P. by the Alexander von Humboldt-Stiftung.
  D.R. was supported, in part, by the Graduiertenkolleg ``Materialien und
Ph\"anomene bei sehr tiefen Temperaturen'' of the DFG.
  J.A.S. acknowledges partial support by the Science and Technology Center
for Superconductivity through NSF Grant no.~91-20000.
  Authors D.R. and J.A.S. also acknowledge additional support from the
Max-Planck-Gesellschaft and the Alexander von Humboldt-Stiftung.

\end{document}